\documentclass{article}
\usepackage{array} 
\usepackage{graphicx}
\usepackage{listings}
\usepackage{float}
\usepackage{framed}
\usepackage{graphicx}
\usepackage[affil-it]{authblk}
\usepackage{amssymb}
\usepackage{authblk}
\usepackage[hyphens]{url}

\usepackage{subfigure}
\usepackage[skip=10pt]{caption}

\usepackage{xcolor}

\newcommand*{\EoE}{\hfill\ensuremath{\square}} % End of Example

\newtheorem{exam}{Example}[section]
\newenvironment{example}{\begin{exam}\rmfamily\upshape}{\end{exam}}

\title{Diagnosing Refactoring Dangers}

 \author[1]{W. Brinksma   \thanks{wouter.brinksma@nhlstenden.com}}
 \author[1]{W. Wernsen    \thanks{we.wernsen@gmail.com}}
\author[1]{E. Verduin    \thanks{evertverduin@gmail.com}}
\author[1]{H. Hilberink  \thanks{herman.hilberink@gmail.com}}
\author[1]{P. de Beer    \thanks{p.debeer@fontys.nl}}
\author[1]{A. Bijlsma
\thanks{lex.bijlsma@ou.nl}}
\author[1]{H. Passier    \thanks{harrie.passier@ou.nl}}

\affil[1]{Open Universiteit\\Faculty of Science\\Department of Computer Science\\Postbus 2960, 6401 DL Heerlen, The Netherlands}

\begin{document}
\maketitle

\lstset{tabsize=4}

\setcounter{section}{0}
\section{Introduction}
\label{sec:introduction}

Software is regularly modified and extended to meet new requirements.
During this process, the underlying structure of the software often deteriorates \cite{Fowler:1999}.
As a result, the cost of extending and adapting the software increases enormously,
a situation also known as technical debt \cite{Kruchten2012}.
To counter this, the structure of software can be improved so that it is easier to understand,  modify and extend, while keeping the same outward behavior. This activity is called Refactoring \cite{Fowler:1999}.

Mens and Trouwé \cite{Mens:2004} described the refactoring process as a number of separate activities:
\begin{enumerate}
  \item Identify where the software should be refactored;
  \item Determine which refactoring(s) should be applied to the identified places;
  \item Guarantee that the applied refactoring preserves behavior;
  \item Apply the refactoring;
  \item Assess the effect of the refactoring on quality characteristics of the software (e.g., complexity, understandability, maintainability) or the process (e.g., productivity, cost, effort);
  \item Maintain the consistency between the refactored program code and other software artifacts (such as documentation, design documents, requirements specifications, tests and so on).
\end{enumerate}

Step three has our special attention:
How can it be guaranteed that a program exhibits the same behavior after a refactoring has been performed?
This is known as behavior preservation \cite{opdyke1992refactoring}.

\subsection{Behavior Preservation}
\label{ch:introduction sec:behavior_preservation}

One way of guaranteeing behavior preservation is to formally prove beforehand the program semantics.
For some languages with formally defined semantics, one can prove that some refactorings preserve the program semantics \cite{proietti1991semantics}.
However, for most complex languages like Java, this is extremely difficult, if not unfeasible \cite{tokuda2001evolving}.

Two other known techniques are `testing' and `refactoring preconditions' \cite{Mens:2004}.
With testing, tests specifying the desired behavior determine whether the specified relationship between input and output has been preserved during the refactoring and after the refactoring has been performed \cite{Fowler:1999}.
Problems with testing include:
\begin{itemize}
  \item Verification takes place after a refactoring has taken place; 
  \item Errors can remain undetected if test cases do not fully cover all the desired behavior;
  \item Failed test cases do indicate that certain desired behavior is no longer being exhibited, but they do not always clearly indicate where the cause can be found in the code, and it is not obvious how the changes can be undone.
  \item Tests can become invalidated by structural changes in the code, which is non-trivial to resolve \cite{Passier:2016}.
\end{itemize}

An example of the last bullet is the Extract Method \cite{Fowler:1999} refactoring as inverse refactoring of Inline Method \cite{Fowler:1999}. Extract Method has many variants. A consequence is that applying Inline Method followed by Extract Method will not necessarily recreate the original program. Depending on which variant is selected, the original or another program is created. If a different program results, the original test may not be applicable because the signature of a method can be different from the original.

Behavior preservation can be achieved by specifying `refactoring preconditions' or `enabling conditions' in advance that must be met before a refactoring is performed \cite{opdyke1992refactoring}.
This technique is applied in many refactoring engines such as Eclipse\footnote{\url{https://eclipseide.org/}} and NetBeans\footnote{\url{https://netbeans.apache.org/}}.
Problems with this technique are:
\begin{itemize}
	\item Refactoring engines may have overly weak and overly strong preconditions with the result that incorrect transformations are allowed \cite{soares:2010} and correct transformations are prevented \cite{mongiovi2018detecting};
	\item The number of ways to establish an enabling condition is often too limited (see explanation after Example \ref{ch:introduction ex:move_method_with_subclass});
	\item It is not always clear why a refactoring is rejected and no remedial action or sufficient alternatives are suggested \cite{debeer2019};
	\item Refactor engines are not open for extension, i.e. it is not possible to add other refactorings.
\end{itemize}
In summary, analyses are often incomplete, little or no insight is given into the reasons why a refactoring is rejected, and no solutions are given for each of the reasons for rejection.
However, programmers prefer that the refactoring engine does not only reject a refactoring application.
Instead, they prefer insight in the reasons of rejection and the possible solutions of how to satisfy the failed preconditions. 
With this insight, they are able to manually fix these problems afterwards~\cite{vakilian2014alternate}.

\subsection{Illustrative Refactoring Examples}
\label{ch:introduction sec:refactoring_example}

The following two examples illustrate the problems mentioned in Section \ref{ch:introduction sec:behavior_preservation}. Example \ref{ch:introduction ex:move_method} presents a situation in which different choices can be made to fulfill a refactoring's precondition. Example \ref{ch:introduction ex:move_method_with_subclass} presents a situation in which a developer should receive a warning of a possible unwanted situation in code. Both examples use Java as the programming language. The end of an example is denoted by the symbol \ensuremath{\square}.

\begin{example}\label{ch:introduction ex:move_method}
% \hspace{0.1pt}

Consider the source code in Figure \ref{ch:introduction fig:example_code_move_method}. Now assume that we wish to apply a Move Method \cite{Fowler:1999} refactoring and move method \lstinline|Source.method| to class \lstinline|Target|.

\begin{figure}[H]
    \centering

\begin{lstlisting}[language=Java]
class Source {

	private int local = 15;

	public void method(Target target) {
		target.doSomething();
		System.out.println("Executing source method with " + local);
	}
}

class Target {

	public void doSomething() {
		System.out.println("Executing target code");
	}
}
\end{lstlisting}
    
    \caption{Example code for Move Method refactoring}
    \label{ch:introduction fig:example_code_move_method}
\end{figure}

\begin{figure}[H]
    \centering

\begin{lstlisting}[language=Java]
class Source {

	private int local;

	public int getLocal() {
		return local;
	}
}

class Target {

	public void doSomething() {
		System.out.println("Executing target code");
	}

	public void method(Source source) {
		doSomething();
		System.out.println("Executing source method with " + source.getLocal());
	}
}
\end{lstlisting}
    
    \caption{Applied Move Method on code from Figure \ref{ch:introduction fig:example_code_move_method}}
    \label{ch:introduction fig:example_code_move_method_refactored}
\end{figure}

A problem that is immediately visible is that attribute \lstinline|local| is not in scope within class \lstinline|Target|. Several solutions are possible for this situation. For instance, class \lstinline|Source| could add a getter to get the value of attribute \lstinline|local|. Attribute \lstinline|local| could also be moved to class \lstinline|Target|, or its visibility could be set to package. If we choose the first option, the resulting code would be that of Figure \ref{ch:introduction fig:example_code_move_method_refactored}. There exist various situations where one of the other options would be preferable. When performing a refactoring, the programmer should be aware of the various possibilities and be able to make an informed choice.
\EoE
\end{example}	

\begin{example}\label{ch:introduction ex:move_method_with_subclass}
Consider the starting situation of Example \ref{ch:introduction ex:move_method} again and assume we want to move method \lstinline|Source.method| to class \lstinline|Target| while class \lstinline|Target| has a subclass \lstinline|Sub|, already having a method \lstinline|method| with parameter \lstinline|source| of type \lstinline|Source|. The additional class is shown in Figure \ref{ch:introduction fig:example_code_move_method_additional}.

\begin{figure}[H]
    \centering

\begin{lstlisting}[language=Java]
class Sub extends Target {

	@Override
	public void method(Source source) {
		System.out.println("Using subclass method");
	}
}
\end{lstlisting}
    
    \caption{Class addition to code from Figure \ref{ch:introduction fig:example_code_move_method}}
    \label{ch:introduction fig:example_code_move_method_additional}
\end{figure}

Note that \lstinline|Sub.method| has the same signature as the new method \lstinline|Target.method| that appeared after applying the Move Method refactoring as shown in Figure \ref{ch:introduction fig:example_code_move_method_refactored}. Since the programmer performing the refactoring may not be aware of the existence of \lstinline|Sub|, there could be a situation of unintended overriding of the \lstinline|Target.method|. A client intending to call \lstinline|Target.method| on an object declared to be of type \lstinline|Target|, will see \lstinline|Sub.method| being executed instead, if at the point of call the object has dynamic type \lstinline|Sub|. Here the refactoring programmer should receive a warning.
\EoE
\end{example}

In Example \ref{ch:introduction ex:move_method}, automatic refactoring in Eclipse\footnote{At least in one version. The behavior of Eclipse tends to change with each new version.} will choose the option to give attribute \lstinline|local| package scope (and provide a warning that this has occurred), without mentioning the other solutions. The problem with subclass \lstinline|Sub| from Example \ref{ch:introduction ex:move_method_with_subclass} is ignored by Eclipse.

In case \lstinline|Source.method| overrides a method in a superclass of \lstinline|Source|, on the other hand, Eclipse refuses to perform the refactoring, and generates a message ``The method invocations to \lstinline|Source.method| cannot be updated, since the original method is used polymorphically''. It is unclear why this refusal occurs because the problem can be solved easily by keeping \lstinline|Source.method| as a shell that redirects all calls to \lstinline|Target.method|.

\subsection{Contributions to the field}
\label{ch:introduction sec:general_research_goal}

The problems mentioned are of interest to professional software developers and to students of computer science. Refactoring code can have associated dangers, i.e. possible behavioral changes of the code (we define this in detail in Section \ref{ch:refactoring_diagnosis_model}). For professional developers it is important to understand exactly which dangers are involved before a refactoring can be applied and how each of these dangers can be solved. 
Depending on this, the decision can be made whether or not to carry out a refactoring and how each of the dangers can be best solved. 
If software engineers have a detailed insight in what the dangers are, they will be better able to judge on that basis whether a refactoring should be carried out and in what form \cite{vakilian2014alternate}.

To teach students to refactor, it is important that they understand the complexities of refactoring and learn how to deal with these complexities.
For this purpose it is important that, for example before starting a refactoring task, an overview is given of the dangers that may arise. 
Various didactic approaches are conceivable for using this insight in programming education. 

In our research and especially in this article, the technology of behavioral preservation by means of static detection of dangers is central.
Rather than preconditions forbidding refactorings, we desire insight in actual dangers and advice on possible remedial actions.
Therefore, we have designed a system for automatic identification of refactoring dangers. We consider topics such as automatic generation of corresponding advices as future work.

In this paper we try to answer the following research question:

\begin{quote}
    How can we conceptually model a refactoring danger analysis in such a way that it makes the dangers posed by a refactoring explicit and complete?
\end{quote}

To answer this research question, we apply Design Science as methodology \cite{wieringa2014design}. Critical evaluation of early implementations \cite{debeer2019, verduin2021, hilberink2021, Wernsen2022} led to the insight that, first of all, the underlying concepts needed to be distinguished with greater precision. Based on this insight we start by designing a conceptual model and build a prototype \cite{brinksma2023} to validate the model in practice. This provides hands-on experience and insight into how refactoring dangers are formed and how they can be avoided. The prototype should provide insight to all dangers that are present and their possible solutions. In addition, the model should make possible later additions of new refactorings.

The remainder of this paper is organized as follows. In Chapter \ref{ch:refactoring_diagnosis_model}, we explain the conceptual model using various examples. Chapter \ref{sec:tool} shows the user interface of the newly created refactoring tool. Next, in Chapter \ref{sec:aspects}, we explain a number of aspects and decisions pertaining to the development of the model and tool. In particular, we show the strong relation between the model and the implementation. Finally, in Chapter \ref{sec:related_work}, we discuss previous work related to our research, and we conclude the paper and define future work in Chapter \ref{sec:conclusion}.

\section{Refactoring Diagnosis Model}
\label{ch:refactoring_diagnosis_model}

To be able to identify and reason about the dangers of refactoring, we need a model of refactoring. This section first introduces this model in the form of a Ubiquitous Language (UL) \cite{evans2004domain}, after which we will explain how dangers can be automatically detected.

\subsection{The model}

The core of our refactoring diagnosis model is that behavior preservation can only be broken if the code of an application is modified. Therefore, we have to analyze the steps of a refactoring for actions that change the code. For now, we assume that the refactoring danger analysis takes place once prior to a refactoring: the user is shown all dangers prior to the refactoring and possible solutions to prevent them. In future, this assumption may be weakened.

\subsubsection*{Step}

Fowler \cite{Fowler:1999} defines a refactoring as a change made to the internal structure of software, maintaining its observable behavior. A refactoring can be divided into a sequence of smaller behavior-preserving transformations called steps. After executing a step, the software should be able to be compiled and run.

\begin{example}
\label{ch:refactoring_diagnosis_model ex:move_method_desc}
Suppose we move a method \lstinline|m| from class \lstinline|A| to \lstinline|B|, assuming a method with the same signature as \lstinline|m| does not yet exist in \lstinline|B|'s namespace. The first step is to examine all program elements used by \lstinline|A.m|, after which we copy \lstinline|A.m| to class \lstinline|B|, resulting in \lstinline|B.m|, and adjust it so it can still access the same elements as \lstinline|A.m|. In the second step, we either change \lstinline|A.m| to call \lstinline|B.m| or replace every call to \lstinline|A.m| with a call to \lstinline|B.m|.
\EoE 
\end{example}

If we break up the steps shown in Example \ref{ch:refactoring_diagnosis_model ex:move_method_desc}, we find that they consist of two smaller entities: the removal of a syntax construct and the addition of one. While a step is behavior preserving, its components are not. We call such components microsteps.

\subsubsection*{Microstep}
Code is modified when a syntax construct is added or removed.
Renaming a syntax element, such as the name of a class or method, can be considered as first removing the element followed by adding a new element. The number of syntax constructs on which both modifications are applicable is made up of all the syntax elements of the programming language. This number is limited too. The combination of a modification (add, remove) and a syntax construct (class, method, attribute, ..., etc.) is what we call a microstep. Adding or removing a syntax construct in a program can have significant impacts on its behavior and functionality.

\begin{example}
\label{ch:refactoring_diagnosis_model ex:remove_method_override}
If class \lstinline|C| contains an override for an inherited method \lstinline|m| and we remove \lstinline|m| without thinking about \lstinline|m| still being present in a superclass of \lstinline|C|, calling method \lstinline|m| on an object of type \lstinline|C| is still possible, but this will execute the inherited method instead of the local one, thus possibly changing the program behavior. Of course, this is only a danger when the original code does not adhere to Liskov's substitution principle.
\EoE 
\end{example}

Because, as seen in Example \ref{ch:refactoring_diagnosis_model ex:remove_method_override}, a microstep can lead to changes in program behavior, a microstep is not behavior-preserving and carries one or more specific potential risks.

\subsubsection*{Potential risk}
Any microstep carries some risks, i.e.it \emph{can} lead to changes in the behavior of the program.

\begin{example} \label{ch:refactoring_diagnosis_model ex:potential_risk}
When we want to add a method \lstinline|m| to a class \lstinline|C|, a matter of overriding could unintentionally happen.
\EoE 
\end{example}

A potential risk is a problem of a microstep that in isolation, i.e. without considering subsequent microsteps and the program code the refactoring is applied to, can change the behavior of a program. For each microstep, all potential risks can theoretically be summed up in advance. It depends on the code context whether a potential risk is an actual risk.

\subsubsection*{Actual risk}
By examining the code context on which the refactoring is applied, it becomes clear if a potential risk is actually present in the code, i.e. one that possibly leads to changed program behavior after the refactoring is completed.

\begin{example}
Following Example \ref{ch:refactoring_diagnosis_model ex:potential_risk}, when we want to add a method \lstinline|m| to a class \lstinline|C|, a matter of overriding happens when class \lstinline|C| has one or more superclasses and one or more of these superclasses owns a method with the same signature, but different semantics, as method \lstinline|m|. This leads to a violation of Liskov's substitution principle.  The same type of risk happens in case one or more subclasses of class \lstinline|C| own a method with the same signature, but different semantics, as method \lstinline|m|. If class \lstinline|C| has no such superclass or subclass, there is no risk of overriding.
\EoE 
\end{example}

An actual risk is a problem of a microstep in the program code the refactoring is applied to, that in isolation, i.e. without considering subsequent microsteps, changes the behavior of the program. Depending on the other microsteps, an actual risk can turn into a danger.

\subsubsection*{Danger}
An actual risk does not always change program behavior eventually. This depends on the complete sequence of microsteps executed as the implementation of the refactoring.

\begin{example} \label{ch:refactoring_diagnosis_model ex:danger_solved}
Classes \lstinline|C| and \lstinline|D| both have a method \lstinline|m|. If we move \lstinline|m| from \lstinline|D| to \lstinline|C| as part of a Move method refactoring, an actual risk of double definition takes place and could turn into a danger. However, when a subsequent microstep removes or renames one of the \lstinline|m| present in \lstinline|C|, the danger is not present any longer.
\EoE 
\end{example}

A danger is defined as an actual risk, caused by a microstep, which is not mitigated by subsequent microsteps. This does not mean that the behavior of the program as it stands will necessarily change. If in Example \ref{ch:introduction ex:move_method_with_subclass} method \lstinline|target.method(source)| is only called in situations where the dynamic type of \lstinline|target| is \lstinline|Target| rather than \lstinline|Sub|, no observable change results. In this case, the danger can not be detected by simply running the program. Static analysis, however, can detect the danger. If the behavior does change, this might lead to an error during runtime.

\subsubsection*{Error}

Improper refactoring can lead to errors being introduced into a codebase. There are two kinds of errors: run- and compile-time.

\begin{example} \label{ch:refactoring_diagnosis_model ex:runtime_error}
When the override method \lstinline|m| is removed from class \lstinline|C| in Example \ref{ch:refactoring_diagnosis_model ex:remove_method_override}, the method might have had some specific effect (i.e. an internal list always being ordered a specific way) needed for objects of type \lstinline|C| to operate correctly. Now with method \lstinline|m| removed, \lstinline|C| will fall back on the superclass implementation of \lstinline|m| which does not have the desired effects. If other parts of \lstinline|C| expect these effects to be present, this might result in runtime errors.
\EoE 
\end{example}

\begin{figure}[htp]
	\begin{center}
		\includegraphics[scale=0.17]{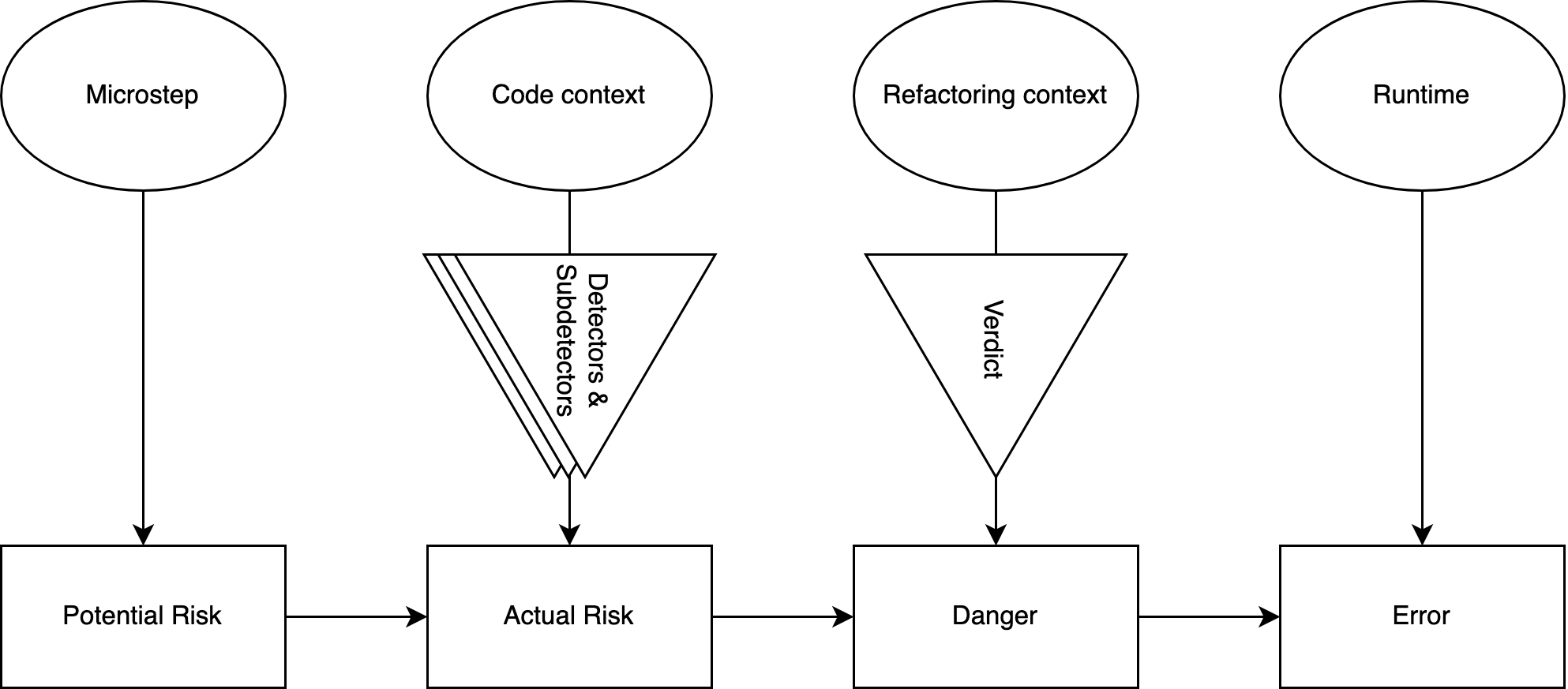}
	\end{center}
	\caption{Type of risks in relation to contexts}
    \label{ch:refactoring_diagnosis_model fig:refactoring_diagnosis_model_risks}
\end{figure}

Figure \ref{ch:refactoring_diagnosis_model fig:refactoring_diagnosis_model_risks} shows all types of risks we distinguish in relation to the types of contexts in which the microsteps are invoked. The last stage shows runtime errors resulting from dangers.

% Sometimes, a danger can become a compile error. 
\noindent Intermediate stages of a refactoring can lead to compile-time errors as well.
\begin{example}
When as part of the Move method refactoring a method has been moved to another class and old references to the method are not yet updated to the new method's location, an error results during compilation.
\end{example}
Both types of errors are out of scope in this article. It should be noted that compilation errors can show up as dangers as well. Furthermore, not every danger will necessarily result in an error. Our model only detects dangers, and not errors.

\subsection{Detection automation}
\label{subsec:detection_automation}

To be able to automatically detect dangers in a refactoring, we realise the previously explained concepts as reusable software entities, resulting in a system consisting of detectors, subdetectors, and a verdict mechanism. The reason why we choose to implement these concepts as separate classes, is that they become reusable in different refactorings, as will be explained in Chapter \ref{sec:aspects}. Below, each of them will be detailed and their interconnections will be explained.

% een detector is een object dat onderzoekt of een potentieel risico actuaal is

\subsubsection*{Detectors}
A detector is a software entity that, given a potential risk and its code context, determines if the potential risk is an actual risk, and if so, at what program locations the actual risk is manifested.

\begin{example} \label{ch:refactoring_diagnosis_model ex:informal_detector}
In case of checking if class \lstinline|C|'s superclasses contain a specific method \lstinline|m|, an informal version of the function looks like Figure \ref{ch:refactoring_diagnosis_model fig:informal_detector} below.
\begin{figure}[H]
    \centering

\begin{lstlisting}
IF C is a class
  AND C has superclasses
  AND m exists in one of these superclasses
THEN list all program locations needed for further analysis
\end{lstlisting}
    
    \caption{Informal detector version}
    \label{ch:refactoring_diagnosis_model fig:informal_detector}
\end{figure}
\EoE 
\end{example}

Example \ref{ch:refactoring_diagnosis_model ex:informal_detector} illustrates a detector. A potential risk can be described using an if-then rule. The if-part sums up the combination of properties that should be present in the code context. The then-part pinpoints all the information needed for further processing. It is very useful if a detector does not produce a boolean result but rather a set of program locations where the actual risk occurs. This will make it possible to inspect these locations to determine whether the risk leads to behavioral changes, and if so, to suggest some remedial change.

Analyzing the potential risks associated with a particular microstep will, in general, take more than one detector. For instance, if we remove a method, we need a detector that will produce all program locations where that method is called. But we will also need a second detector, such as Example \ref{ch:refactoring_diagnosis_model ex:informal_detector} illustrates, that will point to any superclass method that the method to be removed is overriding, because calls to that superclass method will no longer through dynamic binding activate the removed method, thereby possibly changing program behavior.

Detectors are not specific to a particular microstep. For instance, the detector that searches for superclass methods being overridden is also useful when a new method is added: it may be that the new method's name inadvertently leads to it overriding an existing superclass method, again possibly changing program behavior. Hence with every microstep, we associate a number of detectors that look for actual risks, and every one of these detectors is associated with a number of microsteps -- a many-to-many relation.

Each boolean sub-expression in the if-part of a detector can be answered by a function querying the code context on critical locations for the existence of a property. The software entity providing such a function is called a subdetector. Like a detector, a subdetector yields a list of program locations. The boolean result corresponds to that list being empty or not.

% een subdetector is een softwareeenheid die, gegeven de code context en een locatie daarin, een collectie locaties oplevert die voldoen aan een bepaald criterium

\subsubsection*{Subdetectors}
We have defined a subdetector as a software entity that provides a function which, given a set of locations, results in a set of locations that satisfies a specific criterion.

Detectors will pinpoint program locations with actual risks. But these may be thought of as composed of even simpler units, called subdetectors.

\begin{example} \label{ch:refactoring_diagnosis_model ex:informal_subdetectors}
An excerpt of Example \ref{ch:refactoring_diagnosis_model ex:informal_detector} which shows two examples of subdetectors is shown in Figure \ref{ch:refactoring_diagnosis_model fig:informal_subdetectors} below.

\begin{figure}[H]
    \centering

\begin{lstlisting}
AND C has superclasses
AND m exists in one of these superclasses
\end{lstlisting}
    
    \caption{Informal subdetector examples}
    \label{ch:refactoring_diagnosis_model fig:informal_subdetectors}
\end{figure}
\EoE 
\end{example}

As shown in Examples \ref{ch:refactoring_diagnosis_model ex:informal_detector} and \ref{ch:refactoring_diagnosis_model ex:informal_subdetectors}, the detector that searches for superclass methods being overridden may be viewed as the composition of two subdetectors: one that, given a class, produces its ancestors in the inheritance hierarchy; and one that, given a class and a method, produces any occurrence of a method with the same name and signature in that class. We can generalize this by saying that a subdetector always operates on a program location and results in a set of program locations. This also makes it possible that the subdetectors can be successively chained together to form a similar expression such as shown in Example \ref{ch:refactoring_diagnosis_model ex:informal_subdetectors}. A subdetector is not a detector, because a subdetector does not determine whether a potential risk is an actual risk. 

Again there is a many-to-many relationship: subdetectors are not specific to a particular detector, and a detector will normally use several subdetectors. Note that, different from our concept of detector, subdetectors tend to be chained.

To be able to determine for a particular refactoring whether an actual risk is a danger we need a function that analyzes all the actual risks in the refactoring context, i.e. the code context and the set of steps of the refactoring. We call this analysis function a verdict function.

\subsubsection*{Verdict mechanism}

Producing appropriate advice for an entire refactoring, or even a step, necessitates more than just producing the actual risks found by the detectors. It may be that an actual risk associated with one microstep is mitigated by a subsequent one.

\begin{example} \label{ch:refactoring_diagnosis_model ex:verdict_mechanism}
If we perform a change in some method by first adding the new version and then removing the old one, one of the detectors could report an actual risk of a double method definition because at one point during the change, the same method declaration appears twice. However, the actual risk of a double definition does not cause problems because it disappears when the old version is removed. The same holds if we first remove the old version and then add the new one with the actual risk of a missing method definition.
\EoE 
\end{example}

Superfluous warnings, such as in Example \ref{ch:refactoring_diagnosis_model ex:verdict_mechanism}, had better be removed from the advice; this necessitates a `verdict' at a level higher than that of microsteps, at the refactor level. The verdict mechanism is a refactoring-specific operation on actual risks resulting from its detectors, which determines if the actual risk is a danger. This is necessary because detectors consider only one microstep in isolation, while its risks may be obviated by subsequent microsteps in the refactoring. Whereas detectors and subdetectors can be reused over and over again, verdicts so far have not been decomposed, and one must be custom-built for every refactoring. However, this does not exclude possible reuse of verdict code.

\subsection{Model architecture}
\label{ch:refactoring_diagnosis_model sec:model_architecture}

\begin{figure}[H]
	\begin{center}
		\includegraphics[scale=0.2]{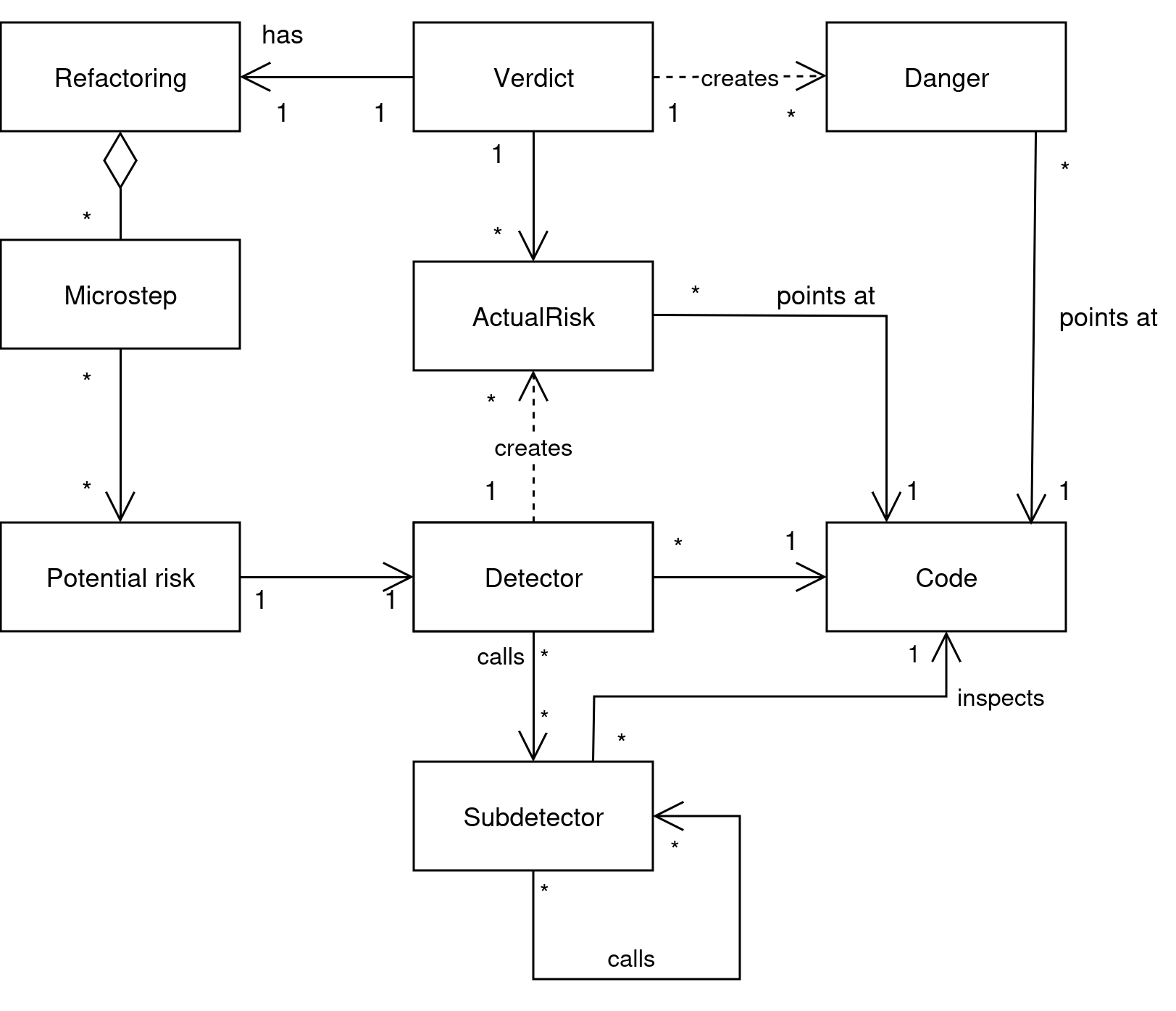}
	\end{center}
	\caption{Domain model of the concepts}
    \label{ch:refactoring_diagnosis_model fig:refactoring_diagnosis_domain_architecture}
\end{figure}

To present the model and its constituent relationships in a visual form, Figure \ref{ch:refactoring_diagnosis_model fig:refactoring_diagnosis_domain_architecture} shows the domain model of the concepts discussed. An implementation of this model is presented in Section \ref{sec:aspects}.

\setcounter{section}{2}
\section{The tool}
\label{sec:tool}
\begin{figure}[H]
    \centering
    \includegraphics[scale=0.23]{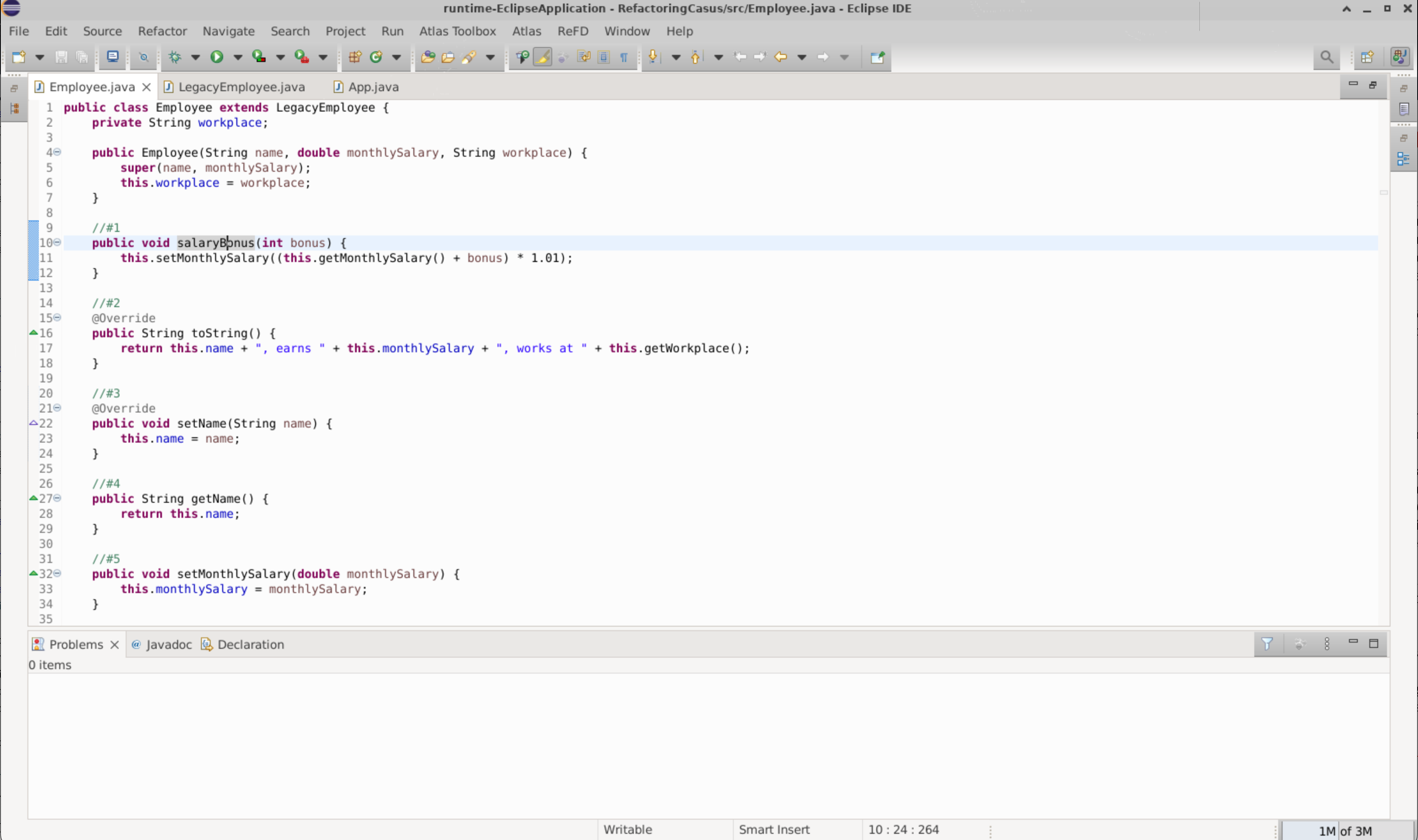}
    \caption{The tool with the refactoring case study example loaded}
    \label{ch:refactoring_diagnosis_plugin fig:tool_normal}
\end{figure}

In this section, we present the prototype Eclipse\footnote{Eclipse version 2023-06 was used} plugin \emph{ReFD}\footnote{ReFD can be accessed through the following GitHub repository:\\\url{https://github.com/NHLStenden-ISAL/ReFD}} \cite{brinksma2023} which shows the execution of the diagnosis process for the Pull Up Method refactoring. Figure \ref{ch:refactoring_diagnosis_plugin fig:tool_normal} shows the normal setup of the tool. A sample project is loaded in the editor. It contains a class \lstinline|Employee| and \lstinline|LegacyEmployee|. \lstinline|Employee| extends \lstinline|LegacyEmployee| and contains some similar methods. We will show how the Pull Up Method refactoring works in the current version of the tool. To do this, we must first select a method to pull up. In Figure \ref{ch:refactoring_diagnosis_plugin fig:tool_normal}, this has already been done, and the method \lstinline|salaryBonus(int)| is selected.

\begin{figure}[H]
    \centering
    \includegraphics[scale=0.27]{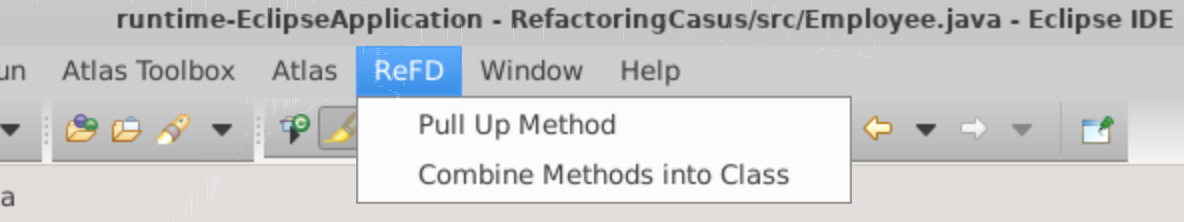}
    \caption{Refactoring selection menu}
    \label{ch:refactoring_diagnosis_plugin fig:tool_menu}
\end{figure}

To initiate the refactoring diagnosis process, the user can select the correct refactoring from the menu as shown in Figure \ref{ch:refactoring_diagnosis_plugin fig:tool_menu}.

\begin{figure}[H]
    \centering
    \includegraphics[scale=0.27]{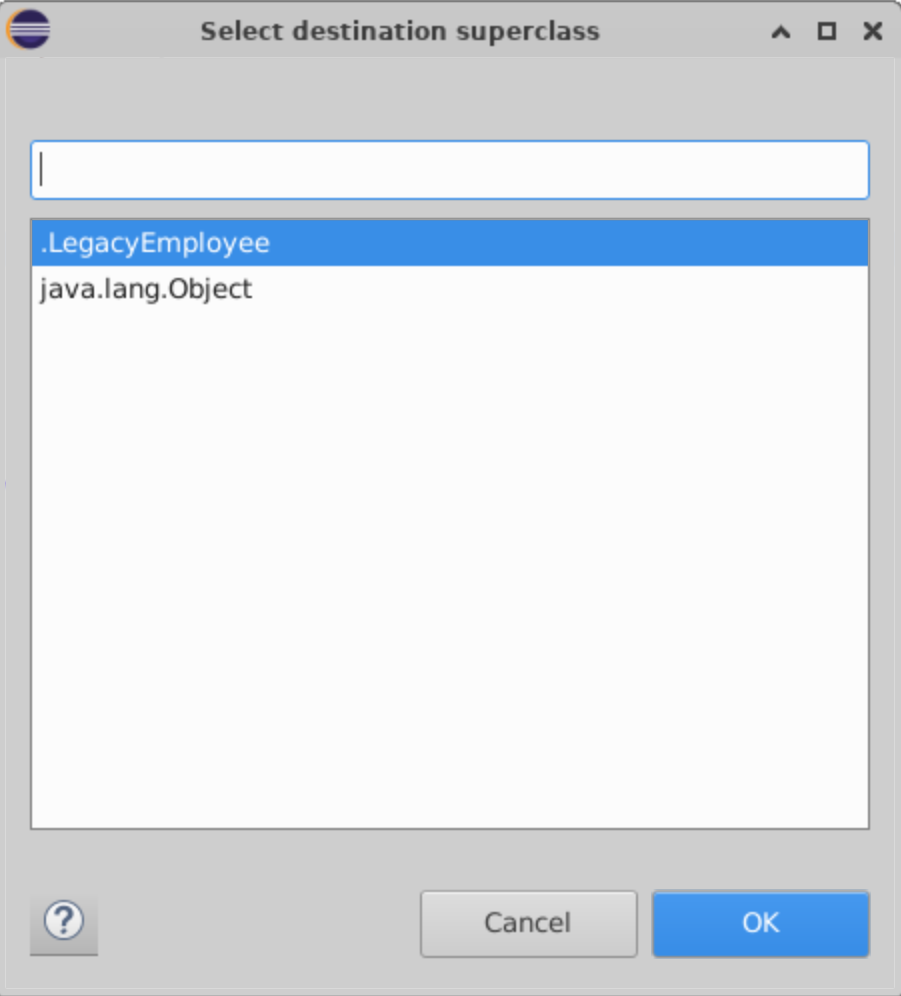}
    \caption{Selecting destination class}
    \label{ch:refactoring_diagnosis_plugin fig:tool_select_destination}
\end{figure}

After selecting the Pull Up Method refactoring entry from the menu, the tool will check if a method is currently selected in the editor. If this is not the case, an error message will be displayed. In this case, because we did select a method, the tool now presents us with a selection screen, as shown in Figure \ref{ch:refactoring_diagnosis_plugin fig:tool_select_destination}. This screen contains all superclasses of the method's enclosing class, in this case superclasses of \lstinline|Employee|. From this list, a superclass must be selected to start the diagnosis process.

\begin{figure}[H]
    \centering
    \includegraphics[scale=0.23]{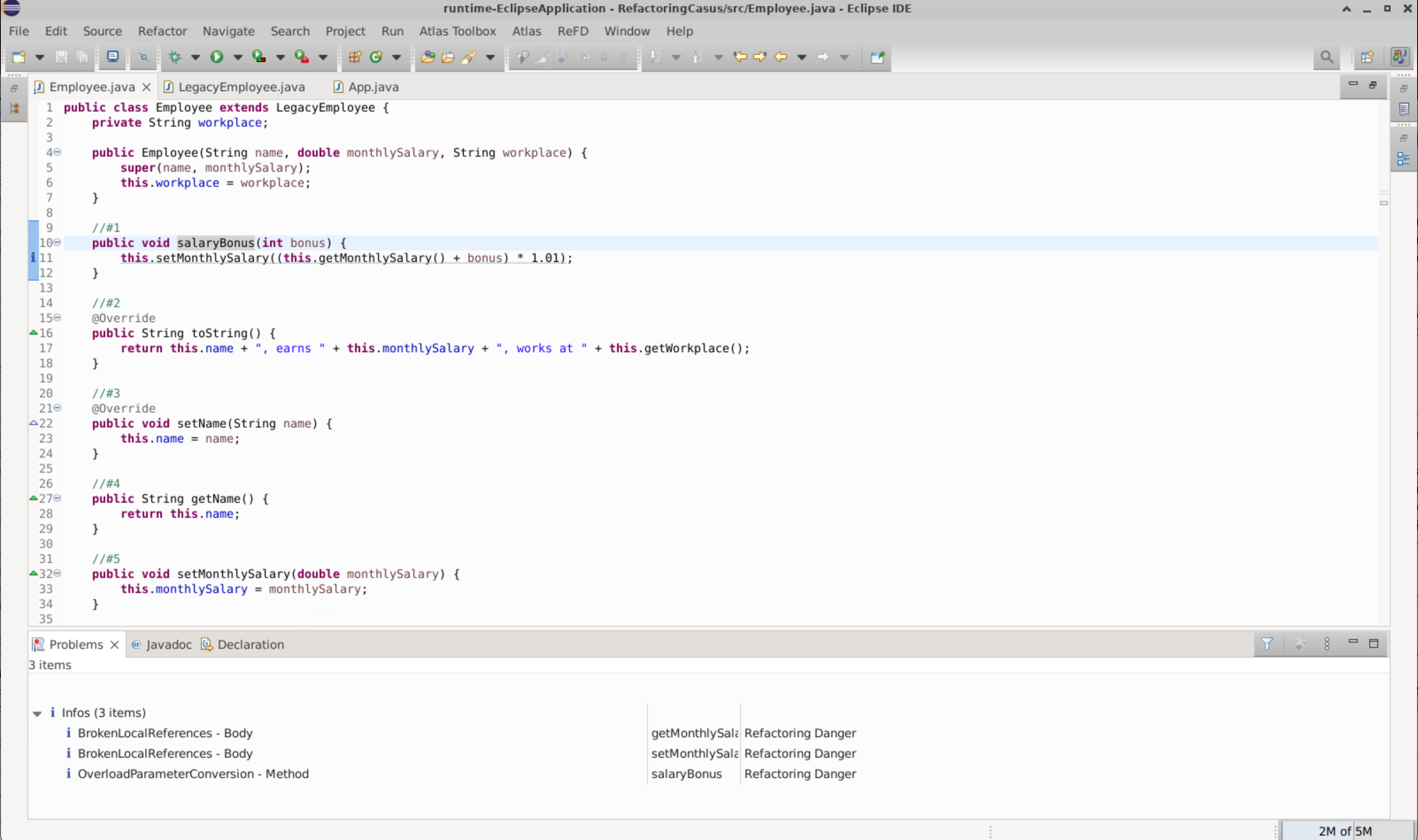}
    \caption{Results from refactoring showing in Problems window}
    \label{ch:refactoring_diagnosis_plugin fig:tool_results}
\end{figure}

We select superclass \lstinline|LegacyEmployee|. Next, the tool will start the analysis process in a different thread to prevent the user interface from lagging during the process. Once the results are in, they are displayed in Eclipse's \emph{Problems} window, as shown in Figure \ref{ch:refactoring_diagnosis_plugin fig:tool_results}. For every danger found, a row containing its description and location in the code will be shown. When the user clicks a row, the tools displays the location in the editor window.

\setcounter{section}{3}
\section{Design aspects of the tool}
\label{sec:aspects}

In this section, we will discuss the design considerations of the tool and give an example of how the refactoring diagnosis model outlined in Section \ref{ch:refactoring_diagnosis_model} can be implemented by way of a query language.

\subsection{Requirements}
As argued above, several microsteps may exhibit the same potential risk. Therefore detectors should not be local to an entity whose purpose is the analysis of a particular microstep. Rather, there should be a library of detectors.

A representation of the program should be available for the detectors to inspect; the result of a detector should then be a set of locations in this representation. An abstract syntax tree and data-flow analysis certainly contain sufficient information, but not all of this is relevant for our purpose. We provide more detail on this subject in Section \ref{ch:refactoring_diagnosis_plugin sec:program_graph}.

Subdetectors represent a single investigation step within the action of a detector, but subdetectors are not specific to a particular risk. Subdetectors should not be local to the code implementing a detector, but form a library as well from which all detectors may select the subdetectors necessary from them, which is shown in Figure \ref{fig:Layers}.

\subsection{Layers}
A formal difference between subdetectors and detectors is that subdetectors do not answer a question about the presence of an actual risk. But as we have determined that a detector can answer this indirectly through the set of code locations produced, detectors and subdetectors both have a set of code locations as input and output. It is, however, practical to keep the libraries for detectors and subdetectors separate, as these are used for different activities: respectively, analyzing a microstep by calling the associated detector for each of its potential risk, and implementing a detector.
\begin{figure}[H]
	\begin{center}
		\includegraphics[scale=0.23]{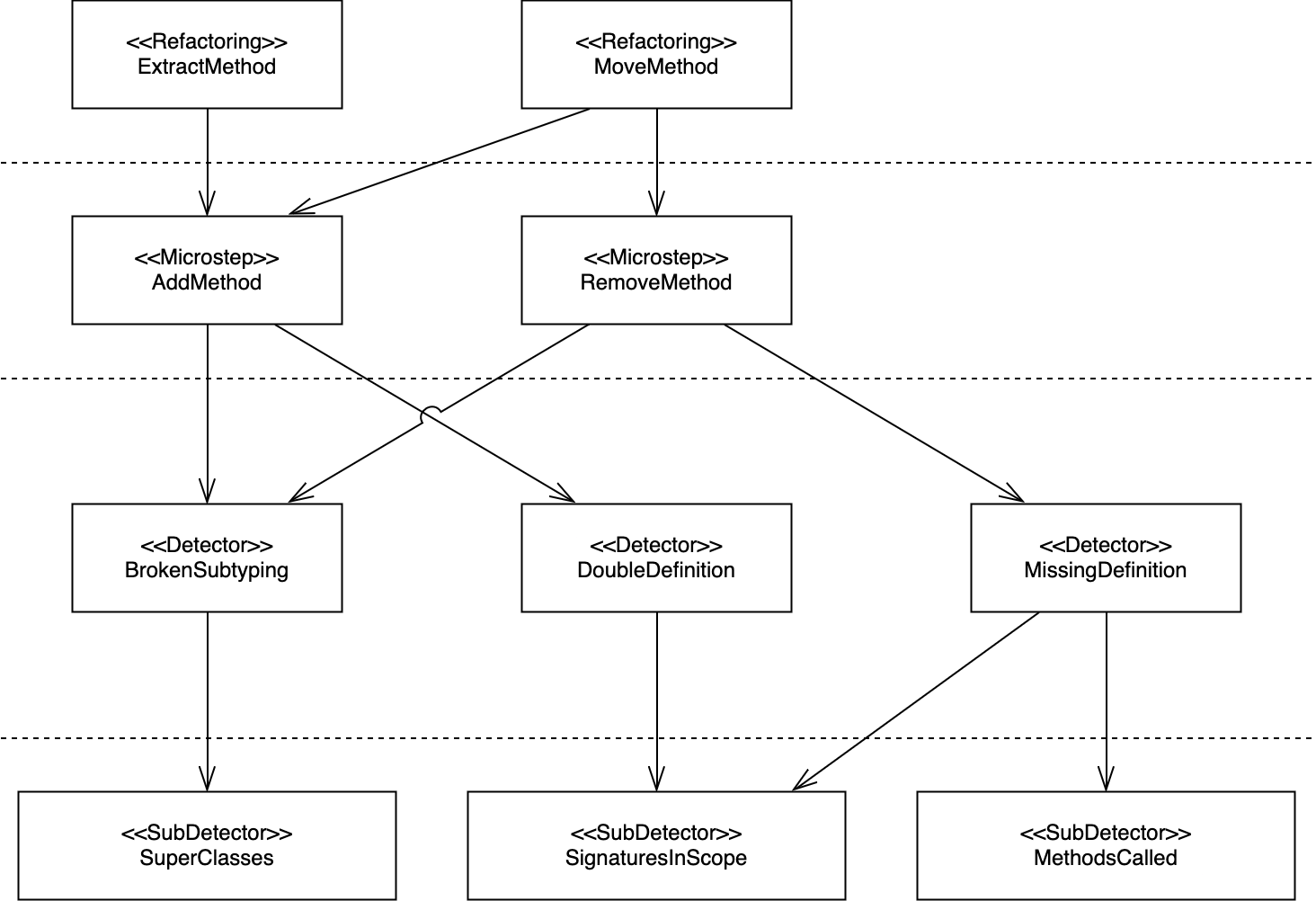}
	\end{center}
	\caption{Layered architecture}
    \label{fig:Layers}
\end{figure}

As microsteps are only studied as phases in the execution of a refactoring, detectors are only called for a given microstep, in order to assess which of its inherent potential dangers are actual dangers in the context of the given program, whereas subdetectors are only called by detectors, these libraries exhibit a \emph{strictly} layered architecture. However, as all communication between these objects proceeds through sets of code locations, the simplified syntax tree representation of the program to be refactored should be globally accessible.

In Figure \ref{fig:Layers} the elements have been tagged with a stereotype, e.g. \lstinline{<<Detector>>}, to point out their role. However, there are still various possibilities for their implementation. For instance, the elements may be realised as classes, and then the stereotypes will be an interface these classes implement. However, as these will be attributeless singleton classes containing just one static function, an implementation as pure functions may be regarded more natural. In that case the stereotypes may become their common function type, or the container in which the functions are kept. Which is more feasible depends on the language chosen for the tool.

\subsection{Plugin implementation}

To explain how the Eclipse plugin works, we consider a simplified version of the implemented Pull Up Method refactoring and analyze its refactoring dangers. We will highlight each important step in the analysis, which will explain the integration and operation of the different elements of the system.

\subsubsection{Program graph}
\label{ch:refactoring_diagnosis_plugin sec:program_graph}

To diagnose an intended refactoring, we need some way of gaining insight into the codebase the refactoring is to be applied to. We do this by converting the codebase into a graph that looks like an Abstract Syntax Tree (AST) but is augmented with a data-flow analysis. To convert the codebase into such a graph, we use an existing tool called Atlas\footnote{\url{https://www.ensoftcorp.com/atlas/}}. The adapter design pattern \cite{Gamma:1995} is applied to the graph Atlas uses, which wraps its elements used by our developed plugin, thereby creating our own version that we will call \emph{program graph}.

Each node in the program graph represents a \emph{program location} that is some location in the codebase, and is of the type \lstinline|ProgramLocation|. These program locations can represent many different elements of a program, such as an entire class or an individual statement. Because these program locations are part of the program graph, they are linked by edges of that graph. Each edge in the program graph represents a \emph{relation} between two program locations, and is of the type \lstinline|Relation|. This method of typing program locations and relations is very coarse-grained. To be able to differentiate between different kinds of program locations and relations, we use a tag. For instance, a program location representing a class will be tagged with a \lstinline|class| tag. This manner of weak typing is used because the graph model is rather abstract, and its only two strong types are \lstinline|ProgramLocation| and \lstinline|Relation|. Atlas handles tagging by using strings, but our software uses specific enumerations to guarantee appropriate tagging. To work with strongly typed program locations, collections must be used. These will be explained in Section \ref{ch:refactoring_diagnosis_plugin sec:streams_and_collections}.

Finally, the program graph contains a facility to query the graph: the \lstinline|GraphQuery|. This too is an adapter that wraps Atlas's query facility called \lstinline|Q|. From one or more starting nodes, \lstinline|GraphQuery| can traverse the program graph.

\begin{example}
\label{ch:refactoring_diagnosis_plugin ex:graph_query_example}
Suppose we have two nodes $m$ and $m'$ representing methods, and a relation of the \emph{override} type defined from $m$ to $m'$. This means method $m$ overrides method $m'$. If a \lstinline|GraphQuery| has $m$ as a starting node, it can query the relations present and traverse the override relation from $m$ to $m'$. \lstinline|GraphQuery| can be thought of to contain a set of program locations as results of a query. The \lstinline|GraphQuery| started with the set $\{m\}$. The subsequent query then altered the set of program locations. This meant that after walking the override relation, the \lstinline|GraphQuery| contained the set $\{m'\}$.
\EoE 
\end{example}

\subsubsection{Representing possibly nonexistent code locations}
\label{ch:refactoring_diagnosis_plugin sec:representing_nonexistent_code_locations}

Sometimes, however, we need to work with a program location that might not exist (yet). As an example, this can happen when adding a new method. To be able to convey its location before the method is added to the codebase or program graph, some medium other than \lstinline|ProgramLocation| must be used because that type of program location always exists in the graph.

To model program locations that already exist, or we are about to introduce in the codebase, we use program location \emph{templates}\footnote{Not to be conflated with C++'s templates.}. A template is an object that contains a description of all defining features of some syntax construct.

\begin{example}
\label{ch:refactoring_diagnosis_plugin ex:location_specification_example}
If we want to describe the method \lstinline|m()| of a class, we use the code in Figure \ref{ch:introduction fig:method_specification_example}.

\begin{figure}[H]
    \centering

\begin{lstlisting}[language=Java]
ClassTemplate parent_template = //creation logic here...
MethodTemplate toString_template =
	new MethodTemplate(
		"m", //method name
		new ArrayList<ParameterSpecification>(), //parameters
		AccessModifier.PUBLIC, //visibility
		false, //static
		false, //abstract
		"void", //return type
		parent_template //parent class
);
\end{lstlisting}
    
    \caption{MethodSpecification usage example for \lstinline|Object.toString()|}
    \label{ch:introduction fig:method_specification_example}
\end{figure}
\EoE 
\end{example}

The template shown in Example \ref{ch:refactoring_diagnosis_plugin ex:location_specification_example} specifies a method that might exist, but this depends on the code context. When we query the graph for a method that matches this template and find a program location, it means the described method exists. If we do not find a location in the graph, the describe method does not exist.

Now that we know how to represent existing program locations, as well as nonexistent ones, we can start to look at how a refactoring is implemented.

\subsubsection{Refactoring}
\label{ch:refactoring_diagnosis_plugin sec:refactoring}

A refactoring is modeled by the abstract class \lstinline|Refactoring|, which provides facilities every refactoring needs. The most important of these facilities is the ability to add microsteps to it.

\begin{example}
\label{ch:refactoring_diagnosis_plugin ex:pull_up_method_example}
In this example, we will discuss the implementation of the Pull Up Method refactoring. Pulling up a method implies that a method should be moved from one class to one of its superclasses, as shown in Figure \ref{ch:introduction fig:pull_up_method_refactoring}.

\begin{figure}[H]
    \centering

\begin{lstlisting}[language=Java]
public class PullUpMethod extends Refactoring {

	private final MethodSpecification target;
    //...
	
	public PullUpMethod(MethodSpecification target, ClassSpecification destination) {
		this.target = target;
        //...
		MethodSpecification newLocation = target.copy();
		newLocation.setEnclosingClass(destination);
		
		addMicrostep(new RelocateMethod(target, newLocation));
	}
	
	//...
	
}
\end{lstlisting}
    
    \caption{PullUpMethod.java}
    \label{ch:introduction fig:pull_up_method_refactoring}
\end{figure}
\EoE 
\end{example}

The class \lstinline{PullUpMethod} shown in Example \ref{ch:refactoring_diagnosis_plugin ex:pull_up_method_example} represents a Pull Up Method refactoring. Refactorings contain microsteps, stored inside an internal list. In this case, the \lstinline{RelocateMethod} microstep is added to this list. We note that the code for the refactoring is rather declarative. It only specifies the constituent elements of the refactoring and no analysis logic needs to be specified here. For \lstinline|PullUpMethod|, this  is done by a \lstinline{RelocateMethod} microstep, which is part of the refactoring analysis decomposition.

\subsubsection{Microstep}
\label{ch:refactoring_diagnosis_plugin sec:microstep}

To decompose refactorings, microsteps are used. A microstep is modeled by the abstract class \lstinline|Microstep|, which provides facilities every microstep needs. Because, as described in Section \ref{ch:refactoring_diagnosis_model}, a microstep carries one or more potential risks, \lstinline|Microstep| provides the ability to add detectors to it. A detector determines if a potential risk constitutes an actual risk.

Additionally, a microstep can also be composed of other microsteps. This is modeled by the abstract class \lstinline|CompositeMicrostep|. The \lstinline|RelocateMethod| microstep we saw in Example \ref{ch:refactoring_diagnosis_plugin ex:pull_up_method_example} is such a microstep, and consists of microsteps \lstinline{AddMethod} and \lstinline{RemoveMethod}, which are part of the greater library of microsteps.

Example \ref{ch:refactoring_diagnosis_plugin ex:microstep_example} shows the \lstinline{AddMethod} microstep carrying multiple potential risks, which are modeled by detectors. They can be added by the provided \lstinline{addPotentialRisk(...)} method, which, similarly to the refactoring, makes microsteps declarative.

\begin{example}
\label{ch:refactoring_diagnosis_plugin ex:microstep_example}
To illustrate how a microstep works, we will focus on the \lstinline{AddMethod} microstep displayed in Figure \ref{ch:introduction fig:add_method_microstep}.

\begin{figure}[H]
    \centering

\begin{lstlisting}[language=Java]
public class AddMethod extends Microstep {
	
	private final MethodSpecification methodToAdd;
	
	public AddMethod(MethodTemplate methodToAdd) {
		this.methodToAdd = methodToAdd;
		
		addPotentialRisk(new DoubleDefinition.Method(methodToAdd));
		addPotentialRisk(new BrokenSubTyping.Method(methodToAdd));
		addPotentialRisk(new CorrespondingSubclassSpecification.Method(methodToAdd));
		addPotentialRisk(new OverloadParameterConversion.Method(methodToAdd));
	}

	//...

}
\end{lstlisting}
    
    \caption{AddMethod.java}
    \label{ch:introduction fig:add_method_microstep}
\end{figure}
\EoE 
\end{example}

\subsubsection{Potential Risk}
\label{ch:refactoring_diagnosis_plugin sec:potential_risk}

A potential risk is modeled by the abstract class \lstinline|PotentialRisk|. Subclasses extending \lstinline|PotentialRisk| contain a name and description of the risk. They provide a concrete detector which can detect if potential risk is an actual risk.

\begin{example}
\label{ch:refactoring_diagnosis_plugin ex:potential_risk_example}
An example of a potential risk can be seen in Example \ref{ch:refactoring_diagnosis_plugin ex:microstep_example}, specifically in the top half. We see a potential risk called \emph{DoubleDefinition.Method}, which is a potential risk of \lstinline|DoubleDefinition| defined for a \lstinline|Method|.
\end{example}

\subsubsection{Detector}
\label{ch:refactoring_diagnosis_plugin sec:detector}

A detector determines if a potential risk is an actual risk. It uses information from the potential risk associated with it. A detector is modeled by the abstract class \lstinline|Detector|. This class is generic and accepts a type parameter that signifies the type of the detector's output, which is a set of program locations.

\begin{example}
\label{ch:refactoring_diagnosis_plugin ex:detector_example}
To explain how a detector works, we show how \lstinline{DoubleDefinition}, the first of the three detectors seen in Example \ref{ch:refactoring_diagnosis_plugin ex:microstep_example}, is implemented in Figure \ref{ch:introduction fig:double_definition_detector}.

\begin{figure}[H]
    \centering

% \begin{lstlisting}[language=Java]
% public final class DoubleDefinition {

% 	public static class Method extends Detector<MethodSet> {

% 		private final MethodSpecification subject;
		
% 		//...
		
% 		@Override
% 		public MethodSet actualRisks() {
% 			return new ProgramComponentsGenerator()
% 					.stream()
% 					.classes()
% 					.classesByName(subject.getEnclosingClass().getClassName())
% 					.methods()
% 					.methodsWithSignature(subject.getMethodName(),
%     									       subject.getParameterTypes()
% 					).collect();
% 		}
		
% 	}
	
% }
% \end{lstlisting}

\begin{lstlisting}[language=Java]
public final class DoubleDefinition {

	public static class Method extends PotentialRisk<MethodSet> {

        private static final String name = "DoubleDefinition.Method";
        private static final String description = "...";
        
        private final MethodTemplate subject;

        //...

        @Override
        public Detector<MethodSet> detectorForRisk() {
            return new DDM_Detector();
        }
		
		public class DDM_Detector extends Detector<MethodSet> {

            //...

            @Override
    		public ActualRisk<MethodSet> actualRisks() {
    			//Collect program locations on which the risk applies...
    		}
      
        }
		
	}
	
}
\end{lstlisting}
    
    \caption{DoubleDefinition.java}
    \label{ch:introduction fig:double_definition_detector}
\end{figure}
\EoE 
\end{example}

Detectors have specific versions depending on the syntax construct they are applied to. These versions are modeled as inner classes that are grouped in an enclosing class representing the type of detector. In this case, we use the name \lstinline|DDM_Detector|, which stands for \emph{Double Definition Method}. This means that the detector shown in Example \ref{ch:refactoring_diagnosis_plugin ex:detector_example} is a \lstinline{DoubleDefinition} detector, specifically for a method. It receives a \lstinline{MethodSpecification} which is the method to be added to a certain context. This detector checks if, were that to happen, a double definition of the method would take place.

Any detector implements the method \lstinline{actualRisks()}, which determines if the potential risk associated with the detector is an actual risk. This method is implemented using several subdetectors, which are explained in Section \ref{ch:refactoring_diagnosis_plugin sec:subdetector}.

\subsubsection{Sets, Streams \& Generators}
\label{ch:refactoring_diagnosis_plugin sec:streams_and_collections}

To be able to ask questions about the codebase a refactoring is to be applied to, we need a system to facilitate this. In Section \ref{ch:refactoring_diagnosis_plugin sec:program_graph}, \lstinline|GraphQuery| might seem to solve this issue. However, in this context, a serious limitation of the graph model is that it is not rich enough to deal with questions specific to a syntax construct, because the graph does not expressly define types such as classes, methods or attributes. Previously, Wernsen \cite{Wernsen2022} created a query object which allowed all queries to apply to any range of types of program location. This would lead to the possibility of a query which wrongly combines multiple types of program location. For example, we could first query the superclasses of a class, after which it queries their return type. This last query is clearly meant for a method, and we need a model that is rich enough to be able to determine which queries are valid, and which are not. We improved upon this by using strongly typed sets of program locations, corresponding streams, and generators. Each of these elements will be explained below briefly.

\subsubsection*{Sets}

To be able to determine which queries can be executed on a program location, we need to know the location's type. Because it is more practical to execute queries on multiple program locations, we enclose the program locations in a set that signifies their type. Examples of these sets are \lstinline|MethodSet|, \lstinline|ClassSet| and \lstinline|FieldSet|.

\begin{example}
In Figure \ref{ch:introduction fig:chaining_example}, we see examples of typed queries. \lstinline|.classes()| provides only classes, which are then filtered by name. Next, the classes are queried for their methods.

\begin{figure}[H]
    \centering

\begin{lstlisting}[language=Java]
//...
.classes()
.classesByName(subject.getEnclosingClass().getClassName())
.methods()
//...
\end{lstlisting}
    
    \caption{Excerpt of queries on different sets}
    \label{ch:introduction fig:chaining_example}
\end{figure}
\EoE 
\end{example}

\subsubsection*{Generators}

Generators are a facility that generate a set of program locations. A generator acts as starting point of a query, by providing the base set of program locations a detector can further refine. Multiple types of generator exist. For example, a \lstinline|ProgramComponentsGenerator| generates a set of all classes in a program. In the same way, \lstinline|InstanceMethodsGenerator| generates all instance methods present in the program.

\subsubsection*{Streams}

Now that we have typed sets of program locations, we can start to apply queries to these sets. From each type of set or generator, a stream of the same type can be created. A stream contains methods that act on the program locations inside the set. For instance, a \lstinline|MethodStream| can filter methods by their name. This filter method accepts a list of program locations, and outputs the filtered program locations. The queries added to the stream are part of a library of what are called subdetectors.

\subsubsection{Subdetector}
\label{ch:refactoring_diagnosis_plugin sec:subdetector}

A subdetector is used by a detector and realises part of a detector's process in determining if a potential risk is an actual risk. Because a subdetector is only part of that process, its responsibilities are not as great. The subdetector itself does not determine the presence of an actual risk. A subdetector merely receives a set of program locations and returns its results as a new set of program locations.

% \hp{deze alinea kan wat korter, veel is eerder uitgelegd.}

A subdetector is modeled by the abstract class \lstinline|Subdetector|. \lstinline|Subdetector| is not fully implemented. Its method \lstinline|applyOn(Set<ProgramLocation>)| needs to be implemented by every concrete subclass.

\begin{example}
\label{ch:refactoring_diagnosis_plugin ex:subdetector_example}
We start to show how a subdetector works by using \lstinline{Overrides} as an example. This subdetector can be seen in Figure \ref{ch:introduction fig:method_subdetectors}.

\begin{figure}[H]
    \centering

\begin{lstlisting}[language=Java]
public final class MethodSubdetectors {
	//...

	public static class Overrides extends Subdetector {

		@Override
		public Set<ProgramLocation> applyOn(Set<ProgramLocation> locations) {
			GraphQuery gq = Graph.query(locations);
   
			return gq.descendantsOn(
					gq.universe()
					  .relations(Tags.Relation.OVERRIDES)
			).locations();
		}

	}

	//...
}
\end{lstlisting}
    
    \caption{MethodSubdetectors.java}
    \label{ch:introduction fig:method_subdetectors}
\end{figure}

The method \lstinline|applyOn| takes a set of program locations. Because its containing class resides in \lstinline|MethodSubdetectors|, we know the elements of the program location set to represent method locations. These method locations are used in a program graph query. The graph query takes all override relations present in the program and finds which of these relations are descendants of the original method locations.

\EoE 
\end{example}

Like detectors, subdetectors are also grouped by type. In Example \ref{ch:refactoring_diagnosis_plugin ex:subdetector_example}, we find one of these groupings to be \lstinline|MethodSubdetectors|. This means that all subdetectors contained in this class expect methods to be supplied to them.

To explain the inner workings of a subdetector, we will look at the more elementary \lstinline{Overrides} subdetector from Example \ref{ch:refactoring_diagnosis_plugin ex:subdetector_example}. The \lstinline{Overrides} subdetector returns the methods that are overridden by the methods supplied to the \lstinline{applyOn(...)} method. It does this by executing a query on the program graph, shown as \lstinline{Graph.query(...)}. The resulting \lstinline{GraphQuery} object can walk the program graph and in this case, traverse relations in the graph tagged with the \lstinline{Tags.Relation.OVERRIDES} tag. To be certain of the result, it then filters the found locations for methods and lastly returns the result.

\subsubsection{Analysis}
\label{ch:refactoring_diagnosis_plugin sec:analysis}

The previous sections explained that parts of the domain are developed in a declarative manner. A refactoring forms a tree with the refactoring at its root, microsteps as internal nodes, and detectors as its leaves. This tree does not contain logic to manage the danger analysis process. Instead, this is implemented in a visitor that visits the refactoring's tree structure. This visitor called \lstinline|DangerAnalyser| walks the tree and aggregates all detector results. However, these results might contain false positives. To remove these, it uses the refactoring's verdict function.

\subsubsection{Verdict}
\label{ch:refactoring_diagnosis_plugin sec:verdict}

To filter false positive results from detectors, a verdict specific to each refactoring is constructed. A base class for such a verdict exists, and is called \lstinline|VerdictFunction|. \lstinline|VerdictFunction| is an abstract class that uses double dispatch in the same way a visitor might. For every detector, \lstinline|VerdictFunction| contains a default visit method that lets every result through. If a detector result should be filtered, its corresponding visit method should be reimplemented. When letting results through, they are labeled with the detector that found them.

In Example \ref{ch:refactoring_diagnosis_plugin ex:verdict_example}, we see that this verdict function has a specific implementation for the detector \lstinline|RemovedConcreteOverride.Method|. This means that when \lstinline|DangerAnalyser| receives results from that detector and passes them to this method, those results are filtered based on its implementation. In this case, we see that it checks if the refactoring pulls the method up to the direct superclass or not. In case we pull up to the direct superclass, we remove the detector's results on line 15 by \lstinline|none(detector)|. In all other cases, we let all results through on line 18 by \lstinline|all(detector)|.

\begin{example}
\label{ch:refactoring_diagnosis_plugin ex:verdict_example}
To explain how the verdict function works, we look at the verdict function for the \lstinline|PullUpMethod| refactoring. The parameters on lines 6 and 8 were removed for brevity.

\begin{figure}[H]
    \centering

\begin{lstlisting}[language=Java]
public class PullUpMethod extends Refactoring {

    private final boolean toDirectSuperclass;
	//...

	@Override
	public VerdictFunction verdictFunction(...) {
		
		return new VerdictFunction(...) {
			//...
			
			@Override
			public void visit(RemovedConcreteOverride.Method detector)
			{
				if (toDirectSuperclass) {
					none(detector);
				}
				else {
					all(detector);
				}
			}
			
			//...
   
		};
		
	}
	
}
\end{lstlisting}
    
    \caption{VerdictFunction for PullUpMethod.java}
    \label{ch:introduction fig:pull_up_method_verdict_function}
\end{figure}
\EoE
\end{example}

\subsubsection{Changing the graph}
\label{ch:refactoring_diagnosis_plugin sec:changing_the_graph}

In some cases, a refactoring is more complex because it adds elements to the code base that previously did not exist. The \emph{Combine Methods into Class} refactoring is such a case because before it moves methods to a class, it first creates that class anew. At first sight, this is rather problematic when working with the program graph, because it only contains code that actually exists at this moment. Therefore, when a method is moved to a new class that does not exist, all queries on the graph will not work correctly.

\begin{example}
\label{ch:refactoring_diagnosis_plugin ex:augment_graph_example}
Suppose we move a method \lstinline|toString()| to a class to be newly created, and next we move another method \lstinline|toString()| to that same class. In this case, the system will not detect a double definition of \lstinline|toString()| at that location because the class does not exist in the graph. Interestingly, even if the class did exist, the system would still not detect a danger. This is because for this to happen, the first \lstinline|toString()| needs to be already present in the class for the double definition to trigger. However, this is not the case.
\EoE 
\end{example}

This can be remedied by having the microsteps of a refactoring augment the program graph. As an example, this would mean that the microstep \lstinline|AddClass| adds a class node to the program graph and adds the correct relations to it. This functionality has been implemented in the microsteps.
\setcounter{section}{4}
\section{Related work}
\label{sec:related_work}

Providing guidance during refactoring to learn this skill could be thought of as less necessary when developers make heavy use of automated refactoring tooling. While we think having a good grasp of refactoring mechanics is always necessary, it is interesting to note that there is evidence for developers being more inclined to refactor their code manually \cite{Murphy2012, Vakilian2012, Miryung2014}. In particular, the paper by Kim et al.~\cite{Miryung2014} highlights a response from a professional developer asking for code understanding tools to aid in manual refactoring.

It may be that such tools are requested because the process of refactoring is not always clear, even from the teaching material. Fowler's \cite{Fowler:1999} famous book provides an example in the form of the mechanics of the Inline Method refactoring. Trailing the list of mechanics, we find a warning that lists several vague complicating factors which, if one encounters them, should be taken as a signal not to execute the refactoring.

Mens and Trouwé \cite{Mens:2004} provide an extensive overview of refactoring. Their paper contains a useful refactoring example and a list of activities of the refactoring process, which start with identifying the location and type of refactorings to be applied, to go on to guaranteeing behavior preservation. To achieve the latter, they list testing, weakening the notion of behavior preservation and formal proofs. Unfortunately, they do not list a diagnosis of dangers associated with refactorings. However, research that is somewhat close to ours does exist. Below, we provide a small overview.

Soares \cite{soares:2010} presents a tool that checks if a refactoring is safe, i.e. if it preserves program behavior. This tool works by generating test cases that should pass for the initial program and the refactored version. Although we think using testing for this purpose has problems, Soares did find numerous bugs in the refactoring engine used by Eclipse, some of which led to changes in behavior preservation that were hard to spot.

Opdyke \cite{opdyke1992refactoring} defined behavior preservation in terms of receiving unchanged output between the original and refactored versions of the program, while providing both with the same input. He also suggests that \emph{preconditions} can be used to check if the program behavior will be the same after a refactoring.

Kniesel and Koch \cite{Kniesel-Composition} introduced a formal model to compose conditional program transformations, i.e. refactorings with preconditions. In this model, a conditional transformation is a pair of a condition and transformation. In particular, a condition is a mapping of a program to a logical value (true or false) and multiple can be strung together by using a conjunction or disjunction. The goal of the conditions is to find if the associated refactoring will be behavior-preserving. Our model, which is compositional as well, focuses on relaying offending program locations instead of only a logical value to gain insight into actual danger advice. In addition, we introduce the concept of a microstep as the fundamental element of a refactoring for which dangers can be found.

Haendler et al.~\cite{haendler1interactive} present a tool that gives the user a program and a description with one or more refactorings to be applied to that program. The tool also provides a UML diagram of the program after successful refactoring and another of the program as-is. Unit tests can be run to assess if the program was refactored correctly and some code quality analysis is also executed. While this is a comprehensive tool, it does not analyze the steps in a refactoring like our proposed tool does and also relies on testing like the tool by Soares.
\setcounter{section}{6}
\section{Conclusion and Future work}
\label{sec:conclusion}

In this paper, we presented a new model to detect refactoring dangers and a small prototype implementation called ReFD. The model detects dangers by splitting a refactoring up into microsteps that have associated potential risks. A detector examines the code context to determine if the potential risk is present in the code, thus becoming an actual risk. The resulting actual risks are evaluated by a verdict mechanism to remove false positives. The final result comprises labeled program locations for which warnings can be given and an advice generated.

The model we proposed in this article resulted from multiple exploratory reserach projects, some of which also resulted in preliminary tools to demonstrate one or more features. The latest of these projects included the prototype implementation presented in this article.

The current refactoring diagnosis model allows for refactoring danger detection to be built in a modular way. We want to further the development of the tool and create a more extensive library of refactorings, building on the detectors and subdetectors already present. This will give us more insight into the way the refactorings can be split up and will help us list the basic components of refactoring advice. In addition, it might also improve the architecture of our model.

However, when building a more extensive library of refactorings, the problem of determining if all potential risks were found during analysis will rear its head again. We must gain more insight into a systematic process to find potential risks for microsteps and a way to determine if our findings are complete.

A more general problem that needs additional research is how a systematic way can be found to transform a refactoring from Fowler's work \cite{Fowler:1999} into a list of microsteps, as this is currently left up to the insightfulness of the researchers.

The system currently only detects dangers, but cannot generate refactoring advice yet. We want to explore the way the current system can be expanded by the addition of refactoring advice generation.

The way the program graph is currently changed is not a very structured process. It could be worthwhile to formalize this aspect of microsteps better so, just like the detection mechanism, a more formal specification of graph changes can be given.

In the future, we want to display these results by way of a dialogue system through which the user can make solution choices depending on the dangers found.

\bibliographystyle{plain}
\bibliography{refactoring_testing}

\section*{Appendix A: List of implemented microsteps and detectors}
\label{ch:refactoring_diagnosis_plugin sec:list_of_implemented_microsteps_and_detectors}

To be able to support both the Pull Up Method and Combine Methods into Class refactorings, several microsteps have to be implemented. Pull Up Method requires us to move a target method from its enclosing class to its destination class. This results in a \emph{MoveMethod} microstep. A MoveMethod is a combination of two microsteps: \emph{AddMethod} to add the method to the destination class, and \emph{RemoveMethod} to remove the method from its original enclosing class.

We note here that the choice to combine AddMethod and RemoveMethod into the microstep MoveMethod is not a trivial decision, and is based on the fact that microsteps have specific potential risks associated with them, as explained in Chapter \ref{ch:refactoring_diagnosis_model}. MoveMethod has a specific potential risk associated with it that is not associated with its constituent microsteps. This will be explained in the following sections.

The Combine Methods into Class refactoring also moves methods from one class to another, which means we can reuse our previously found MoveMethod microstep. The methods are moved to a newly created class, which means we need a microstep to create a new class: \emph{AddClass}. This analysis leads us to the following list of microsteps: AddMethod (Table \ref{ch:refactoring_diagnosis_plugin table:add_method_potential_risks_detectors}), RemoveMethod (Table \ref{ch:refactoring_diagnosis_plugin table:remove_method_potential_risks_detectors}), MoveMethod (Table \ref{ch:refactoring_diagnosis_plugin table:move_method_potential_risks_detectors}), and AddClass (Table \ref{ch:refactoring_diagnosis_plugin table:add_class_potential_risks_detectors}).

The microsteps found correspond to the work of Verduin \cite{verduin2021}, in which he also provides a preliminary description of hazards associated with those microsteps. These hazards can be thought of as potential risks because they have the same form. Working from these potential risks as a base, we added multiple additional potential risks. Below, we will list for each microstep the associated potential risks and their corresponding detector. Each combination of potential risk and detector is labeled with two letters corresponding to the list of microsteps above, and an index number.

\begin{table}[!htbp]
	\caption{Potential risks for AddMethod microstep and associated detectors}
	\label{ch:refactoring_diagnosis_plugin table:add_method_potential_risks_detectors}
	\small
	\begin{tabular}{ l l l }
		\cline {1-1} \cline{3-3} %\cline{5-5}
		\emph{Label} & & \emph{Detector \& Potential Risk}\\ % & & \emph{Potential Risk}\\
        \cline {1-1} \cline{3-3} %\cline{5-5}
		AM-1 && \underline{DoubleDefinition} - Method to add is already defined in context of target location\\[3mm]
		AM-2 && \underline{BrokenSubTyping} - Method to add is defined in superclass of context as well,\\
             && thus overriding that method in context\\[3mm]
		AM-3 && \underline{CorrespondingSubclassSpecification} - When adding a method, and a method with\\
             && the same signature exists in one or more subclasses, its specification might not\\
             && correspond to these existing methods\\[3mm]
		AM-4 && \underline{OverloadParameterConversion} - Method to add overloads existing method but\\
             && parameter types precede in automatic type conversion rules, thus leading to previous\\
             && calls to the existing method now calling the new method unintentionally\\
		\cline {1-1} \cline{3-3} %\cline{5-5}
	\end{tabular}
    \end{table}

\begin{table}[!htbp]
	\caption{Potential risks for RemoveMethod microstep and associated detectors}
	\label{ch:refactoring_diagnosis_plugin table:remove_method_potential_risks_detectors}
	\small
	\begin{tabular}{ l l l }
		\cline {1-1} \cline{3-3} %\cline{5-5}
		\emph{Label} & & \emph{Detector \& Potential Risk}\\ % & & \emph{Potential Risk}\\
        \cline {1-1} \cline{3-3} %\cline{5-5}
		RM-1 && \underline{MissingDefinition} - Method still called after removal\\[3mm]
		RM-2 && \underline{RemovedConcreteOverride} - Override method is removed, thus defaulting to super\\
             && implementation\\[3mm]
		RM-3 && \underline{LostSpecification} - When method is removed, methods overriding it will not\\
             && have a specification anymore to adhere to\\[3mm]
		RM-4 && \underline{MissingSuperImplementation} - Method to be removed is concrete and not all of the\\
             && containing class's subclasses have an override implementation, thus leading to possible\\
             && missing implementations\\[3mm]
        RM-5 && \underline{MissingAbstractImplementation} - Method to be removed implements abstract method\\
             && in concrete class, while its implementation is mandatory\\
		\cline {1-1} \cline{3-3} %\cline{5-5}
	\end{tabular}
    \end{table}

\begin{table}[!htbp]
	\caption{Potential risks for MoveMethod microstep and associated detectors (The MoveMethod microstep includes all potential risks of AddMethod and RemoveMethod, which are not listed here.)}
	\label{ch:refactoring_diagnosis_plugin table:move_method_potential_risks_detectors}
	\small
	\begin{tabular}{ l l l }
		\cline {1-1} \cline{3-3} %\cline{5-5}
		\emph{Label} & & \emph{Detector \& Potential Risk}\\ % & & \emph{Potential Risk}\\
        \cline {1-1} \cline{3-3} %\cline{5-5}
		MM-1 && \underline{BrokenLocalReferences} - Method body contains references to elements from the\\
             && local context (current class and superclasses), such as fields and methods\\
		\cline {1-1} \cline{3-3} %\cline{5-5}
	\end{tabular}
    \end{table}

\begin{table}[!htbp]
	\caption{Potential risks for AddClass microstep and associated detectors}
	\label{ch:refactoring_diagnosis_plugin table:add_class_potential_risks_detectors}
	\small
	\begin{tabular}{ l l l }
		\cline {1-1} \cline{3-3} %\cline{5-5}
		\emph{Label} & & \emph{Detector \& Potential Risk}\\ % & & \emph{Potential Risk}\\
        \cline {1-1} \cline{3-3} %\cline{5-5}
		AC-1 && \underline{DoubleDefinition} - Class to add is already defined in context of target location\\
		\cline {1-1} \cline{3-3} %\cline{5-5}
	\end{tabular}
    \end{table}

\end{document}